\documentclass[12pt]{article}
\pdfoutput=1
\usepackage{jhep-mod}
\title{How to Wick rotate generic curved spacetime}
\author{\large Matt Visser}
\emailAdd{matt.visser@sms.vuw.ac.nz}
\affiliation{School of Mathematics and Statistics,
Victoria University of Wellington; \\
PO Box 600, Wellington 6140, New Zealand.}
\abstract{\\
It is an article of folklore that the collection of ideas
identified as Euclidean quantum gravity may be derived from ordinary
Lorentzian signature gravity by the procedure of Wick rotation.  
This note will attempt to shed some light on this relatively ill--understood
procedure.  I argue that it proves inappropriate and unhelpful to regard Wick rotation in terms of a complex deformation of the time coordinate.  Rather, Wick rotation can more usefully be viewed as a complex deformation of the
spacetime metric. This simple reformulation of the Wick rotation procedure, while it leaves flat space physics unaffected, has profound implications for quantum gravity.

\bigskip\noindent
O{\sc riginal date:} March 1991. This old pre-arXiv essay was originally written for the 1991 Gravity Research Foundation essay contest, while I was affiliated with the Physics Department of Washington University in Saint Louis.  
For many years a PDF copy of the essay was available at the GRF website, but many of the older essays have now been removed from that website.  Apart from purely cosmetic issues, this arXiv upload is an accurate reflection of that 1991 essay.

\bigskip\noindent
K{\sc eywords:} Wick rotation; Euclidean QFT; Euclidean quantum gravity. 

\bigskip\noindent
P{\sc acs:} 04.20.-q, 04.20.Cv, 04.60.+n 
}
\begin{document}
\maketitle
\def\D{{\cal D}} 
\section{Introduction}

A considerable amount of interest is currently focussed on the complicated
collection of ideas known as Euclidean quantum gravity.  Now, it is an
article of folklore that Euclidean quantum gravity may be derived from
ordinary Lorentzian signature gravity by the procedure of Wick rotation.
While Wick rotation is a well understood process in flat spacetime, its
application to curved spacetimes is more problematic and less
straightforward than one might have reason to hope. It is the purpose of
this note to shed some light on this ill--understood procedure.  

In a quantum field theory defined on flat Minkowski space, Wick rotation is
physically justified by appealing to the causality constraints, as embodied
in Feynman's ``$i\epsilon$'' prescription \cite{Itzykson-Zuber,Bjorken-Drell}. 
After Wick rotating from
Minkowski space to Euclidean space these causality constraints survive in a
modified form as ``Osterwalder--Schrader positivity'' (OS positivity) of the
Euclidean version of the field theory \cite{Glimm-Jaffe}. On the other hand,
in the usual formulation of the transition from Lorentzian signature gravity
to Euclidean signature quantum gravity the effects of the causal structure
of the Lorentzian signature theory seem to disappear completely as no
analogue of OS positivity is manifest \cite{Hawking}. This circumstance is
profoundly disturbing, and suggests that something crucial is missing in the
naive analysis.

\clearpage

A related observation, indicating that the process of Wick rotating a
general spacetime is less well-understood than one might suppose, is that
the naive prescription ``$t\to -it$'' does not in general yield a real
Euclidean signature metric.  For example: when acting on a stationary
spacetime, this prescription (interpreted as a ``generalized coordinate
transformation'') does succeed in making the time--time and space--space
components of the metric both real and positive \cite{conventions}. However,
even for this relatively simple case the ``$t\to -it$'' prescription will
render the time--space components of the metric pure imaginary \cite{Morris}.
The problem becomes even more acute in non--stationary spacetimes where even
the time--time and space--space components of the metric may become complex.
For instance, a particularly transparent example is provided by ordinary
de~Sitter space \cite{Redmount}. In the comoving coordinates corresponding to a spatially flat $k=0$ slicing, de~Sitter space may be described by the metric
\begin{equation}
ds^2 = -dt^2 + H^{-2} \;e^{Ht}\; (dr^2 + r^2(d\theta^2 +\sin^2\theta d\phi^2)).
\end{equation}
Naive analytic continuation yields
\begin{equation}
ds^2 = +dt^2 + H^{-2}\;e^{-iHt}\;(dr^2 + r^2(d\theta^2+ \sin^2\theta d\phi^2)).
\end{equation}
While certainly entertaining, this manifestly complex analytically continued metric is hardly
the hoped for ``Euclideanization'' of de~Sitter space.  Worse, the act of
analytic continuation depends on the coordinate system employed.  The metric
of de~Sitter space may equally well be written down in terms of the comoving coordinates corresponding to a  positive spatial curvature $k=+1$ slicing, so that
\begin{equation}
ds^2 = -dt^2 
     + H^{-2}\;{\rm cosh}^2 Ht \;\left({dr^2\over 1-r^2}
     + r^2(d\theta^2 + \sin^2\theta d\phi^2)\right). 
\end{equation}
Naive analytic continuation in this case yields a metric that is manifestly
Euclidean:
\begin{equation}
ds^2 = +dt^2 
     + H^{-2}\;\cos^2 Ht \; \left({dr^2\over 1-r^2} 
     + r^2(d\theta^2 + \sin^2\theta d\phi^2)\right). 
\end{equation}
In fact this analytically continued metric has a natural interpretation as
the canonical metric on a $S^4$ hypersphere of radius $H^{-1}$, and we see
that the often quoted result that \underbar{the} Euclideanization of
de~Sitter space is the 4--sphere is a coordinate dependent result. For the comoving coordinates corresponding to a negative spatial curvature $k=-1$ slicing, the metric of de~Sitter space is
\begin{equation}
ds^2 = -dt^2 
     + H^{-2}\;{\rm sinh}^2 Ht \; \left({dr^2\over 1+r^2} 
     + r^2(d\theta^2 + \sin^2\theta d\phi^2)\right). 
\end{equation}
In this case, naive analytic continuation yields a metric that while real is certainly not Euclidean, the original Lorentzian metric of signature (1,3) being analytically continued to another Lorentzian metric of signature (3,1) given by
\begin{equation}
ds^2 = +dt^2 
     - H^{-2} \; \sin^2 Ht \; \left({dr^2\over 1+r^2} 
     + r^2(d\theta^2 + \sin^2\theta d\phi^2)\right). 
\end{equation}

This explicit coordinate dependence of the naive Wick rotation procedure should inspire a deep and abiding feeling of unease --- clearly something critical is amiss. The situation is further complicated by observing that a general curved spacetime may not even possess a global coordinate patch (indeed de~Sitter space itself possesses this feature). In such a case, the
differentiable structure of the manifold is given in terms of an atlas of
coordinate charts together with a set of transition functions connecting
overlapping charts \cite{Bishop-Goldberg,Hawking-Ellis}. It is far from clear
how to interpret a naive ``$t\to -it$'' prescription on such a structure.

Such concerns may be dealt with by reinterpreting Wick rotation in terms of
an analytic continuation of the metric, leaving the coordinate charts
invariant.  The (reinterpreted) Wick rotation is then a method for connecting
a real Minkowski signature metric with an associated real Euclidean
signature metric while leaving the differentiable structure of the manifold
fixed.  In the case of flat space this prescription will precisely reproduce
the known results, but this viewpoint has the great advantage that it may
meaningfully be extended to general curved spacetimes.

\section{Wick rotation in flat spacetime}
\subsection{The ``$i\epsilon$'' prescription}

Recall the elementary result that causality in Minkowski space is
responsible for the usual ``$i\epsilon$'' prescription for quantum field
propagators \cite{Itzykson-Zuber,Bjorken-Drell}. For instance, the Feynman
propagator for a scalar field is [$P=(E,p)$]:
\begin{equation}
\Delta_F(P) = {i\over E^2 -p^2 -m^2 +i\epsilon}\;.
\end{equation} 
Now, in the complex energy plane the poles in this propagator
occur at energies $E=\pm\sqrt{m^2+p^2-i\epsilon}$, so that when performing the usual
Wick rotation, $E\to iE$, the contour does not pass over the poles. [The
contour $(-\infty,+\infty)$ is deformed to $(-i\infty,+i\infty)$.]  In terms
of the ``rotated'' energy variable the Euclidean propagator is:
\begin{equation}
\Delta_E(P) = {-i\over E^2 +p^2 +m^2}\;.
\end{equation} 
 Since energy and time are Fourier conjugates of each other, it is usual to
rephrase this deformation of the contour of integration over energy in terms
of an analytic continuation of the time variable ``$t\to -it$''.
Unfortunately, as we have just seen, this simple procedure does not have a
consistent generalization when considering generic curved manifolds.

\subsection{Deforming the Minkowski metric}

To obtain an interpretation of Wick rotation that does generalize nicely to
curved manifolds we note that the Minkowski space Feynman propagator may be
recast as:
\begin{equation}
\Delta_F(P) = {i\over E^2(1+i\epsilon) -p^2 -m^2}\;.
\end{equation}
The poles in this propagator occur at
\begin{equation}
E=\pm\sqrt{(m^2+p^2)/(1+i\epsilon)}=\pm\sqrt{m^2+p^2-i\epsilon}. 
\end{equation}
 Here we have used the fact that $\epsilon$ is infinitesimal and that
$m^2+p^2$ is guaranteed to be positive. When the propagator is written in
this form the $i\epsilon$ prescription has a natural interpretation in terms
of a complex ``not quite Minkowski'' metric.  Specifically, let us define
\begin{equation}
\eta_\epsilon = {\rm diag} (-1-i\epsilon, +1, +1, +1).
\end{equation}
In terms of the four momentum and this ``not quite Minkowski'' metric the
Feynman propagator can be compactly written as:
\begin{equation}
\Delta_F(P) = {i\over - \eta_\epsilon(P,P) - m^2}\;.
\end{equation} 
 The prescription for Wick rotation is now also clear. Keeping the real part
of $\epsilon$ positive, analytically continue this metric from $\epsilon=0$
to $\epsilon=+2i$.  (We particularly wish to avoid passing through the point
$\epsilon=+i$ since the metric $\eta_\epsilon$ is degenerate at that point.)
Then $\eta_{(\epsilon=0)} = \eta_L$, while $\eta_{(\epsilon=+2i)} = \eta_E$.
We also point out that $\sqrt{-\det(\eta_\epsilon)} = \sqrt{1+i\epsilon}$,
so that $\sqrt{-\det(\eta_L)}=+1$ and $\sqrt{-\det(\eta_E)} = +i =
+i\sqrt{\det(\eta_E)}$. Thus this prescription for continuing the metric ---
not the coordinates --- is completely equivalent to the usual Wick rotation
in flat spacetime. Moreover, this procedure will now be shown to generalize nicely to curved spacetime.
As a preliminary step to this generalization, we introduce the constant
time--like vector $V=(1,0,0,0)$.  Then we may compactly write
\begin{equation}
\eta_\epsilon = \eta_L + i\epsilon { V\otimes V\over \eta_L(V,V)}\;.
\end{equation}

\section{Wick rotation in generic spacetimes}
\subsection{Deforming Lorentzian--signature metrics}
To apply these ideas to generic curved spacetimes we shall need to invoke
some additional technical machinery. It is well known that a manifold admits a
time-orientable Lorentzian metric if and only if it also admits an
everywhere non-vanishing timelike vector field \cite{Hawking-Ellis}. 

We now construct the ``not quite Lorentzian'' metric:
\begin{equation}
g_\epsilon = g_L + i\epsilon\; {V\otimes V\over g_L(V,V)}\;.
\label{metric}
\end{equation}
 Note that in the flat space limit $g_L\to\eta_L$, and $V\to(1,0,0,0)$, so that one recovers the result of the previous section. Further, by construction,
$g_E = g_{(\epsilon=+2i)}$ is a globally defined (positive definite) Euclidean
metric on the original manifold.  Neither $V$ nor $g_E$ are in any sense
unique, a circumstance that is less than pleasing, but unavoidable.  (In the
case of flat space, $V$ could be taken to be any constant timelike vector.
Although $V$ is not then unique, in the limit $\epsilon\to+2i$, it proves to
be the case that $\eta_E$ is independent of $V$.) 

To see in what sense the metric (\ref{metric}) embodies the curved space version of the ``$i\epsilon$'' prescription we argue as follows:  
For the case of a fixed background geometry one is still provided with a fixed light--cone structure so that one may unambiguously impose the constraint that $[\hat\phi(x),\hat\phi(y)]=0$
whenever the events $x$ and $y$ are spacelike separated.  Now by invoking the (strong) equivalence principle we know that \underbar{locally} spacetime always ``looks like'' a piece of Minkowski space.  More precisely, consider modes whose wavelengths $\lambda$ are much smaller than the local ``radius of curvature'' $\rho= (R_{\alpha\beta\gamma\delta} R^{\alpha\beta\gamma\delta})^{-1/4}$, and whose frequency $\nu$ is much larger than $c/\rho$. For such high frequency/short wavelength modes propagation through spacetime is in all important respects equivalent to propagation through flat space up to terms of order $\lambda/\rho$ or $\nu c/\rho$. It is in this sense that the ``not quite Lorentzian'' metric of equation (\ref{metric}) implies and is implied by curved space causality in the high frequency/short wavelength limit.

\subsection{Quantum gravity: A functional integration over geometries}

The situation is considerably more complex when one seeks to quantize the gravitational field itself. There is no generally agreed upon method for doing this. I have argued elsewhere \cite{Visser} in favour of the primacy of the Lorentzian path integral approach. Adopting this approach for the time being we write the physical partition function as
\begin{equation}
Z_L = \int \D g_L \exp\left( -i\int R(g_L) \sqrt{-g_L}\right).    
\end{equation}
Proceeding formally, using the definition $g_\epsilon\equiv g_L +i\epsilon \{V\otimes V/ g_L(V,V)\}$, we now extend this $Z_L$ to an analytically continued object $Z(\epsilon)$:
\begin{equation}
Z(\epsilon) 
  \equiv \int \D g_L \D V 
    \exp\left( -i\int R(g_\epsilon) \sqrt{-g_\epsilon} \right) 
  \equiv \int \D g_\epsilon 
    \exp\left( -i\int R(g_\epsilon) \sqrt{-g_\epsilon} \right).
\end{equation}
Note that as $\epsilon \to 0$, $Z(\epsilon)\to Z_L$ up to a multiplicative constant. On the other hand for $\epsilon\to +2i$ we may define a Wick rotated Euclidean partition function by
\begin{equation}
Z_E = \lim_{\epsilon\to+2i} Z(\epsilon) 
    \propto \int \D g_E \exp\left( +\int R(g_E) \sqrt{g_E}\right).
\end{equation}
The convergence of this Euclidean path integral is of course problematic due to the well known conformal instability --- for a nice discussion see Mazur and Mottola \cite{Mazur-Mottola}. 

The point I wish to emphasize in this note concerns the range of integration of the $\D g_E$ functional integration over Euclidean metrics. Within the framework espoused in this note the situation is clear: The original functional integration was over Lorentzian metrics defined on Lorentzian manifolds --- and the only Euclidean metrics that can be obtained by Wick rotation from a Lorentzian metric are defined on precisely those Euclidean manifolds that are compatible with the existence of a Lorentzian structure.

It is thus my claim that the $\D g_E$ functional integral should \underbar{not} be taken to be a functional integral over \underbar{all} Euclidean manifolds --- rather it is my claim that the range of integration is restricted for physical reasons to include only those Euclidean manifolds compatible with the existence of a Lorentzian structure.

\section{Discussion}
It is worthwhile to examine carefully the logic of the argument presented in this essay:
\begin{itemize}
\item[{\bf1:}] By specific example I have shown that a naive ``$t\to-it$'' prescription for defining Wick rotation is mathematically inconsistent and physically unsupportable.
\item[{\bf2:}] The improved Wick rotation prescription that I advocate has the virtues of being mathematically well defined and physically reasonable in that it reproduces sensible flat space results. In particular, this version of Wick rotation does not disturb the topological structure of the manifold --- if the manifold has a Lorentzian metric initially then the Wick rotated Euclidean metric will still be compatible with the existence of a Lorentzian metric. 
\item[{\bf3:}] If we treat the Lorentzian path integral as paramount we see that Wick rotation does not lead to arbitrary Euclidean metrics, and the associated Euclidean functional integral over Euclidean metrics is over a topologically restricted class of Euclidean metrics.
\end{itemize}
This restriction of the class of Euclidean manifolds to be included in the Euclidean functional integral is of great importance to understanding quantum gravity. One way of avoiding point (3) --- though in no way diminishing the strength of points (1) and (2) --- is to define the Euclidean path integral \underbar{by fiat} to include all Euclidean manifolds. If such a course is adopted it is extremely difficult to see how a Wick rotation back to Lorentzian signature might be meaningfully defined --- If manifolds that do not admit a Lorentzian metric are included in the Euclidean path integral then they will not suddenly acquire a Lorentzian interpretation through Wick rotation.

At another level, I would argue that defining the Euclidean path integral by fiat is physically unreasonable in that it violates Ockham's razor --- we live in a Lorentzian signature universe --- we believe that the path integral formalism gives a good description of quantum physics --- arbitrarily extending the range of integration of the path integral to include manifolds that do not admit a Lorentzian structure is in no way supported by physical experiment, nor does it appear to add insight to the mathematical problems encountered.

In summary: The naive ``$t\to-it$'' prescription for Wick rotation is physically and mathematically diseased when applied to questions of quantum gravity. There is a fundamental issue of paramount importance hiding in the seemingly innocuous question ``What is Wick rotation?''.

\acknowledgments

I wish to thank Michael S. Morris for a penetrating question that provoked my interest in these topics. I also wish to thank Ian H. Redmount for his interest and comments.  This research was supported (1991) by the U.S. Department of Energy.

\bigskip
\noindent
(Current research supported by the Marsden Fund, administered by the Royal Society of New Zealand.)

\clearpage

\vspace{-10pt}
\section*{Note added (2017):}
Gary Gibbons has kindly pointed out to to me the considerably earlier 1977 article by Candelas and Raine~\cite{Candelas:1977}, which explored similar issues. Additionally, some early 1977 discussion of the conformal mode problem in Euclidean quantum gravity can be found in reference~\cite{Gibbons:1977}. 
In 1987--1988 some related work by Ivashchuk appeared in the Russian-language literature~\cite{Ivashchuk:1987,Ivashchuk:1988}.

\clearpage
If I were writing this essay today, I would add a discussion of Jeff Greensite's ideas regarding dynamical signature change~\cite{Greensite:1992, Greensite:1993, Greensite:1994, Greensite:1995}, the ``complex lapse'' approach of Sean Hayward~\cite{Hayward:1995}, related work by Vladimir Ivashchuk~\cite{Ivashchuk:1997}, and the ``real way'' advocated by Fernando Barbero~\cite{Barbero:1995}.   In the causal dynamical triangulation (CDT) formulation of quantum gravity~\cite{Ambjorn:2006},  restricting the configuration space to Euclidean simplicial manifolds that are Wick rotations of Lorentzian simplicial manifolds seems essential to keeping the functional integral under control.  Other related work over the past decade includes~\cite{Kosyakov:2007, Girelli:2008, White:2008, Helleland:2015, Samuel:2015}. 
A recent MSc thesis related to this topic is that by Finnian Gray~\cite{Gray}. 

\renewcommand{\refname}{Additional References}


\begin{thebibliography}{69}
\bibitem{Itzykson-Zuber}
C. Itzykson and J.-B. Zuber, {\sl Quantum Field
Theory}, (McGraw--Hill, New York, 1980).

\bibitem{Bjorken-Drell}
J. D. Bjorken and S. D. Drell, {\sl Relativistic 
Quantum Fields}, (McGraw--Hill, New York, 1965).

\bibitem{Glimm-Jaffe}
J. Glimm and A. Jaffe, {\sl Quantum Physics: A
Functional Integral Point of View}, (Springer-Verlag, New York, 1987).

\bibitem{Hawking}
An overview of the status of formal developments in
Euclidean quantum gravity may be gleaned from:\hfil\break
S. W. Hawking and W. Israel, {\sl General Relativity: An Einstein Centenary
Survey}, (Cambridge University Press, Cambridge, 1979);\hfil\break
S. W. Hawking and W. Israel, {\sl 300 Years of Gravitation}, (Cambridge
University Press, Cambridge, 1987).

\bibitem{conventions}
We choose our metric conventions such that the flat
space Minkowski metric is $\eta_L = {\rm diag}(-1,+1,+1,+1)$. The flat space
Euclidean metric is taken to be positive definite, $\eta_E = {\rm
diag}(+1,+1,+1,+1)$.

\bibitem{Morris}
This point has been forcefully made by Michael S. Morris,
(private communication).

\bibitem{Redmount}
I wish to thank Ian H. Redmount for bringing this
example to my attention.

\bibitem{Bishop-Goldberg}
R. L. Bishop and S. I. Goldberg, {\sl Tensor
Analysis on Manifolds}, \\
(Dover, New York, 1980).

\bibitem{Hawking-Ellis}
S. W. Hawking and G. F. R. Ellis, {\sl The large
scale structure of space-time}, \\
(Cambridge University Press, Cambridge,
1977).

\bibitem{Visser}
M. Visser, ``Wormholes, Baby Universes and Causality'',\\
 Physical  Review {\bf D41} (1990) 1116.  \; doi:10.1103/PhysRevD.41.1116

\bibitem{Mazur-Mottola}
P. O. Mazur and E. Mottola, 
``The gravitational measure, solution of the conformal factor problem, and stability of the ground state of quantum gravity'',  
\\
Nuclear Physics {\bf B341} (1990) 187.\; doi:10.1016/0550-3213(90)90268-I

\end{thebibliography}

\begin{thebibliography}{69}
\setcounter{NAT@ctr}{11} 

\bibitem{Candelas:1977}
P.~Candelas and D.~J.~Raine,
  ``Feynman propagator in curved space-time'',\\
  Phys.\ Rev.\ D {\bf 15} (1977) 1494.
  doi:10.1103/PhysRevD.15.1494
  
  \bibitem{Gibbons:1977}
  G.~W.~Gibbons,
  ``The Einstein action of Riemannian metrics and its relation to quantum gravity and thermodynamics'',
  Phys.\ Lett.\ A {\bf 61} (1977) 3.
  doi:10.1016/0375-9601(77)90244-4
  
 \bibitem{Ivashchuk:1987}
 V.D. Ivashchuk, ``Regularization by $\epsilon$-metric: I'', 
 \\
 Izvestiya Akademii Nauk Moldavskoi SSR. \\
 Ser. Fiziko-tekhnicheskih i matematicheskih nauk. No. 3, p. 8-17 (1987).  [In Russian.]

 \bibitem{Ivashchuk:1988}
V.D. Ivashchuk. Regularization by $\epsilon$-metric: II. The limit $\epsilon= 0^+$'', 
\\
Izvestiya Akademii Nauk Moldavskoi SSR. \\
Ser. Fiziko-tekhnicheskih i matematicheskih nauk. No. 1, p. 10-20 (1988). 
 [In Russian.]
  
  

\bibitem{Greensite:1992}
  J.~Greensite,
  ``Stability and signature in quantum gravity''.
  
  \bibitem{Greensite:1993}
  J.~Greensite,
  ``Dynamical origin of the Lorentzian signature of space-time'',\\
  Phys.\ Lett.\ B {\bf 300} (1993) 34
  doi:10.1016/0370-2693(93)90744-3
  [gr-qc/9210008].

\bibitem{Carlini:1993up}
  A.~Carlini and J.~Greensite,
  ``Why is space-time Lorentzian?'',
  Phys.\ Rev.\ D {\bf 49} (1994) 866
  doi:10.1103/PhysRevD.49.866
  [gr-qc/9308012].

\bibitem{Greensite:1994}
  J.~Greensite,
  ``Quantum mechanics of space-time signature'',\\
  Acta Phys.\ Polon.\ B {\bf 25} (1994) 5.

\bibitem{Greensite:1995}
A.~Carlini and J.~Greensite,\\
 ``Square root actions, metric signature, and the path integral of quantum gravity'',\\
  Phys.\ Rev.\ D {\bf 52} (1995) 6947
  doi:10.1103/PhysRevD.52.6947
  [gr-qc/9502023].


\bibitem{Hayward:1995}
  S.~A.~Hayward,
  ``Complex lapse, complex action and path integrals'',\\
  Phys.\ Rev.\ D {\bf 53} (1996) 5664
  doi:10.1103/PhysRevD.53.5664
  [gr-qc/9511007].
  
\bibitem{Ivashchuk:1997}  
V.D. Ivashchuk, ``Wick rotation, regularization of propagators by a complex metric and multidimensional cosmology'',
Grav. Cosmol. {\bf3} (1997) 8-16, arXiv:gr-qc/9705008 

  
   \bibitem{Barbero:1995}
  J.~F.~Barbero G.,
  ``From Euclidean to Lorentzian general relativity: The real way'',
  Phys.\ Rev.\ D {\bf 54} (1996) 1492
  doi:10.1103/PhysRevD.54.1492
  [gr-qc/9605066].

  
  
 \bibitem{Ambjorn:2006} 
  J.~Ambjorn, J.~Jurkiewicz and R.~Loll,\\
  ``Quantum gravity, or the art of building spacetime'',\\
  In *Oriti, D. (ed.): Approaches to quantum gravity* 341-359
  [hep-th/0604212].
  
 
\bibitem{Kosyakov:2007}
  B.~P.~Kosyakov,
  ``Black holes: Interfacing the classical and the quantum'',\\
  Found.\ Phys.\  {\bf 38} (2008) 678
  doi:10.1007/s10701-008-9227-z
  [arXiv:0707.2749 [gr-qc]].


\bibitem{Girelli:2008}
  F.~Girelli, S.~Liberati and L.~Sindoni,\\
  ``Emergence of Lorentzian signature and scalar gravity'',\\
  Phys.\ Rev.\ D {\bf 79} (2009) 044019
  doi:10.1103/PhysRevD.79.044019\\{}
  [arXiv:0806.4239 [gr-qc]].

\bibitem{White:2008}
  A.~White, S.~Weinfurtner and M.~Visser,\\
  ``Signature change events: A Challenge for quantum gravity?'',\\
  Class.\ Quant.\ Grav.\  {\bf 27} (2010) 045007
  doi:10.1088/0264-9381/27/4/045007\\{}
  [arXiv:0812.3744 [gr-qc]].


  
  \bibitem{Helleland:2015}
  C.~Helleland and S.~Hervik,\\
  ``A Wick-rotatable metric is purely electric'',
  arXiv:1504.01244 [math-ph].

\bibitem{Samuel:2015}
  J.~Samuel,\\
  ``Wick Rotation in the Tangent Space'',
  Class.\ Quant.\ Grav.\  {\bf 33} (2016)  015006
  doi:10.1088/0264-9381/33/1/015006
  [arXiv:1510.07365 [gr-qc]].

\bibitem{Gray}
Finnian Gray,\\
``Black hole radiation, greybody factors, and generalised Wick rotation'',\\
MSc thesis, 2016, Victoria University of Wellington. 
{\sf  http://hdl.handle.net/10063/5148} 


\end{thebibliography}
\end{document}